\documentclass[fleqn,usenatbib]{mnras}

\usepackage[T1]{fontenc}

\DeclareRobustCommand{\VAN}[3]{#2}
\let\VANthebibliography\thebibliography
\def\thebibliography{\DeclareRobustCommand{\VAN}[3]{##3}\VANthebibliography}

\usepackage{tablefootnote}
\usepackage{listings} 

\usepackage{threeparttable}

\usepackage{graphicx}	
\usepackage{amsmath}	
\usepackage{amssymb}	

\usepackage{newtxtext,newtxmath}

\title[Mock Surveys]{Mock Galaxy Surveys for HST and JWST from the IllustrisTNG Simulations}

\author[G. F. Snyder et al.]{
Gregory F. Snyder$^{1}$\thanks{email: gsnyder@stsci.edu}, 
Theodore Pe\~{n}a$^{1,2}$,
L. Y. Aaron Yung$^{3}$, 
Caitlin Rose$^{4}$,
Jeyhan Kartaltepe$^{4}$,
Harry Ferguson$^{1}$\\
$^{1}$Space Telescope Science Institute, 3700 San Martin Dr, Baltimore, MD 21218\\
$^{2}$University of Wisconsin-Madison, Department of Astronomy, 2535 Sterling Hall, 475 N. Charter Street, Madison, WI 53706\\
$^{3}$Astrophysics Science Division, NASA Goddard Space Flight Center, 8800 Greenbelt Rd, Greenbelt, MD 20771, USA\\
$^{4}$Rochester Institute of Technology, 84 Lomb Memorial Drive,
Rochester, NY 14623
}
\date{Accepted XXX. Received YYY; in original form ZZZ}

\pubyear{2022}

\begin{document}
\label{firstpage}
\pagerange{\pageref{firstpage}--\pageref{lastpage}}
\maketitle

\begin{abstract}

We present and analyze a series of synthetic galaxy survey fields based on the IllustrisTNG Simulation suite. With the Illustris public data release and JupyterLab service, we generated a set of twelve lightcone catalogs covering areas from 5 to 365 square arcminutes, similar to several JWST Cycle 1 programs, including JADES, CEERS, PRIMER, and NGDEEP. From these catalogs, we queried the public API to generate simple mock images in a series of broadband filters used by JWST-NIRCam and the Hubble Space Telescope cameras. This procedure generates wide-area simulated mosaic images that can support investigating the predicted evolution of galaxies alongside real data. Using these mocks, we demonstrate a few simple science cases, including morphological evolution and close pair selection. We publicly release the catalogs and mock images through MAST, along with the code used to generate these projects, so that the astrophysics community can make use of these products in their scientific analyses of JWST deep field observations.

\end{abstract}





\section{Introduction}

With rapidly expanding computing power, it has become possible to make detailed theoretical predictions of entire galaxy populations, together with their internal structures, across cosmic time \citep[e.g.,][]{Vogelsberger2014a,Schaye2014,Dubois2014}. Such simulations also provide an opportunity to forward model their outputs into synthetic data products for comparing directly with observations \citep[e.g.,][]{Overzier2012,Trayford2015,Trayford2017,Torrey2015,Snyder2015,Kaviraj2017,Dickinson2018,Yung2019,Snyder2019,Rodriguez-Gomez2019,Bottrell2019,Huertas-Company2019,Ferreira2020,Pena2021}. In particular, the spatial scales resolved by recent large galaxy formation simulations are approximately one kiloparsec or smaller, roughly matched to the resolving power of the Hubble Space Telescope (HST) and JWST for galaxies observed at $z \gtrsim 1$. Thus, such simulations are ideal candidates for creating synthetic data products to analyze simultaneously with real data.

Such products can be used to make progress on several important questions about galaxy formation in combination with forthcoming JWST data. Two examples that we highlight here are:
\begin{itemize}
\item How does the merger rate evolve in distant galaxies?
\item When and how do bulges and disks form in massive galaxies?
\end{itemize}

With HST survey data, recent studies have found that the galaxy merger rate increases with redshift up to and beyond $z > 3$ \citep{Man2016,Mantha2018,Duncan2019,Ferreira2020}. These inferences rely on having simulation comparison data in order to measure merger observability times \citep[e.g.,][]{lotz08,Snyder2017} to be used in calculating merger event rates from merger fractions. By resolving the rest-frame optical light of galaxies beyond cosmic noon, JWST survey data will yield merger candidates using both pair-based and morphology-based methods. To use those data to constrain the merger rate, we will also need appropriate merger observability times derived from simulations.

In this paper, we create and study a set of mock extragalactic surveys from the publicly available IllustrisTNG simulations \citep{Weinberger2017,Naiman2018,Springel2018,Marinacci2018,Nelson2018,Pillepich2018,Pillepich2018_tngmethods,Pillepich2019_tng50,Nelson2019_tng50,Nelson2019}\footnote{https://tng-project.org}. The primary purpose of this paper is to present the mock data products so that others can use them in their investigations. Section~\ref{s:methods} describes the methods we used to create these mock surveys, and Section~\ref{s:analysis} presents a few scientific analyses demonstrating the use of these products.

\section{Methods} \label{s:methods}

In this section, we describe how we created mock galaxy surveys using the lightcone technique from the publicly available IllustrisTNG simulations, and how we created synthetic pristine HST and JWST mock images based on those lightcones. The resulting data products are made publicly available as a MAST HLSP, with DOI \href{https://doi.org/10.17909/T98385}{https://doi.org/10.17909/T98385} and URL \href{https://archive.stsci.edu/hlsp/illustris}{https://archive.stsci.edu/hlsp/illustris}. The source code used to generate these data products is available at \href{https://github.com/gsnyder206/mock-surveys/releases/tag/v1.0.0}{https://github.com/gsnyder206/mock-surveys/releases/tag/v1.0.0}.

\begin{figure*}
\begin{center}
\includegraphics[width=7.5in]{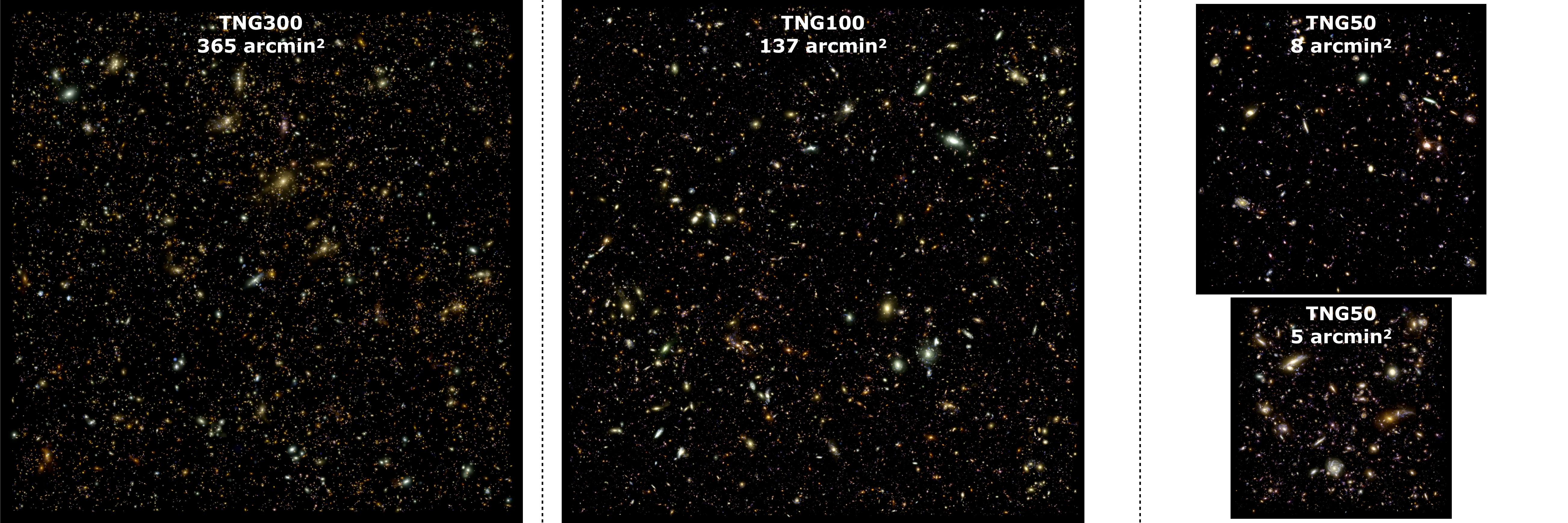}
\caption{Color image renderings of the mock survey fields presented in this paper. These renderings include no PSF or noise effects. The RGB channels include data from three filters each:  B = HST ACS F435W, F606W, and F814W; G = JWST NIRCam F115W, F150W, F200W; and R= F277W, F356W, and F444W. \label{fig:mockrenders}}
\end{center}
\end{figure*}

\begin{table*}
\begin{threeparttable}
\centering
\caption{Properties of the lightcone catalogs and mock images from IllustrisTNG simulations. The ``Images'' column indicates whether or not we created a set of mock images from the associated catalog at the time of this writing.}
\label{tab:lightcones}
\begin{tabular}{ccccccccc}
ID & Simulation & m,n & Direction & Area & redshift limits & g mag. limit& Images & Comparable programs \\
 & & & & (arcmin$^2$) & & \\
\hline
TNG50-12-11-xyz   &   TNG50-1  & 12,11 & xyz & 5 & $0.1 < z < 12$ & 32 & yes & \\
TNG50-12-11-yxz   &   TNG50-1  & 12,11 & yxz & 5 & $0.1 < z < 12$ & 32 & no & NGDEEP \\
TNG50-12-11-zyx   &   TNG50-1  & 12,11 & zyx & 5 & $0.1 < z < 12$ & 32 & no & \\
\hline
TNG50-11-10-xyz   &   TNG50-1  & 11,10 & xyz & 8 & $0.1 < z < 12$ & 32 & yes & \\
TNG50-11-10-yxz   &   TNG50-1  & 11,10 & yxz & 8 & $0.1 < z < 12$ & 32 & yes & JADES, CEERS, NGDEEP\\
TNG50-11-10-zyx   &   TNG50-1  & 11,10 & zyx & 8 & $0.1 < z < 12$ & 32 & yes & \\
\hline
TNG100-7-6-xyz   &   TNG100-1  & 7,6 & xyz & 137 & $0.1 < z < 8.8$ & 30 & yes & \\
TNG100-7-6-yxz   &   TNG100-1  & 7,6 & yxz & 137 & $0.1 < z < 8.8$ & 30 & yes & JADES, CEERS, PRIMER\\
TNG100-7-6-zyx   &   TNG100-1  & 7,6 & zyx & 137 & $0.1 < z < 8.8$ & 30 & no & \\
\hline
TNG300-6-5-xyz   &   TNG300-1  & 6,5 & xyz & 365 & $0.1 < z < 8.8$ & 30 & yes$^*$ & \\
TNG300-6-5-yxz   &   TNG300-1  & 6,5 & yxz & 365 & $0.1 < z < 8.8$ & 30 & no & PRIMER, COSMOS-Web \\
TNG300-6-5-zyx   &   TNG300-1  & 6,5 & zyx & 365 & $0.1 < z < 8.8$ & 30 & no & \\
\end{tabular}
\begin{tablenotes}
\item[*] For the TNG300 mock surveys, the example images only contain sources with $g < 25$.
\end{tablenotes}
\end{threeparttable}
\end{table*}

\subsection{Lightcone Catalogs}

The lightcone technique is a common and effective way of making direct contact between theory and observations of distant galaxies. It has been used regularly for empirical models \citep[e.g.,][]{Behroozi2019,Behroozi2020,Drakos2022}, semi-analytic models \citep[e.g.,][]{Overzier2012,Bernyk2016,Yung2022}, as well as hydrodynamical models \citep[e.g.,][]{Kaviraj2017,Snyder2017,Pena2021}. With the lightcone technique, hydrodynamical models allow us to trace both the large scale structures as well as detailed internal galaxy morphologies via mock images.

We follow the procedure outlined by \citet{Snyder2017} to generate mock sky survey catalogs out of the IllustrisTNG catalogs hosted in the public data release \citep{Nelson2019}. In short, this procedure uses methods from \citep{Kitzbichler2007} to convert from three-dimensional positions of galaxies at multiple time steps in a periodic cubic volume into a three-dimensional lightcone reflecting the evolution of galaxies as observed by a hypothetical observer at the current time. For efficiency, we took advantage of the Illustris JupyterLab service to quickly parse and convert the IllustrisTNG galaxy catalogs into lightcones, without having to download all the subhalo information locally. We focused on lightcone geometries that make best use of the volumes simulated in the IllustrisTNG suite and also reasonably match important extragalactic survey volumes. Table~\ref{tab:lightcones} presents the set of lightcones that we created for this purpose. 

We use the parameters $(m, n)$ to identify three specific geometries relative to the origin of the simulations at z=0, with up to three simple realizations by permuting the x, y, and z directions in the procedure. These $(m,n)$ pairs are $(6,5)$ for the widest lightcones we analyze here ($365$ arcmin$^2$), $(7,6)$ for the middle size ($137$ arcmin$^2$), and $(11,10)$ and $(12,11)$ for the narrower lightcones ($8$ and $5$ arcmin$^2$). The result is twelve total lightcone catalogs:  four sets with three lightcones each. The lightcones span a range of properties comparable to several JWST Cycle 1 programs highlighted in Table~\ref{tab:lightcones}  \citep[e.g.,][]{2017jwst.prop.1345F,2021jwst.prop.1837D,2021jwst.prop.1727K,2021jwst.prop.2079F}.

In order to get the most volume out of these simulations with the lightcone technique, we selected parameters such that some of the co-moving volume of each simulation is repeated a few times throughout its redshift range. In other words, a given subhalo may appear multiple times at widely separated redshifts, each time with its evolutionarily correct appearance and position.

\subsection{Mock Image Mosaics}

For a subset of the IllustrisTNG lightcones, we also generated pristine mock images in a variety of filters. We identify the lightcones for which we created images in Table~\ref{tab:lightcones}, and we show three examples in Figure~\ref{fig:mockrenders}.

We first create blank square mock image arrays in FITS format using the Astropy software  \citep{astropy2013, astropy2018}. These blank images are centered at right ascension and declination values of $0.0, 0.0$, and we create an appropriate world coordinate system (WCS) with principal axes aligned with the RA and Dec directions. 

We then iterate over the lightcone catalog created in the previous section, identify each subhalo from the IllustrisTNG catalogs, and create an image in one of many filters using the public IllustrisTNG web application programming interface \citep[API,][]{Nelson2019}. These images of stellar light do not have any dust modeling performed on them. The returned object is an HDF5-format bytestream, which we first save to a virtual HDF5 file-like object using the Python IO library. Next, we read the data array using the h5py library, and convert it to an Astropy FITS object in memory. See the released code for this procedure to obtain individual subhalo images from the Illustris API and convert them to FITS format. 

We then add these cutouts at their appropriate locations in the mock image array, using the Cutout2D utility in Astropy. We continue until all subhalos have been added, where possible. There are occasional failures, which we identify in the catalogs. Failure modes include sources that overlap with the image edges, as well as occasional API connection issues. The resulting images are pristine mock extragalactic fields from the IllustrisTNG simulations. The flux images have 0.03 arcsec (ACS, NIRCam) or 0.06 arcsec (WFC3) pixels and are saved in units of nanoJanskies.

We generate mock images in the following filters or quantities, which are available as quantities from the Illustris visualization API: 
\begin{itemize}
    \item HST ACS F435W, F606W, and F814W
    \item HST WFC3 F125W, F140W, and F160W
    \item JWST NIRCam F090W, F115W, F150W, F200W, F277W, F356W, and F444W
    \item Stellar mass, in units of $M_*$
    \item Star formation rate, in units of $M_*$ per year
\end{itemize}

The mock images provided by the API use FSPS stellar population models \citep{conroy09_fsps,conroy10_fsps,python-fsps} with Padova isochrones, MILES stellar library, and Chabrier initial mass function. Star particles are rendered using the standard SPH spline technique with adaptive sizes \citep{Nelson2019}.

\begin{figure}
\begin{center}
\includegraphics[width=3.0in]{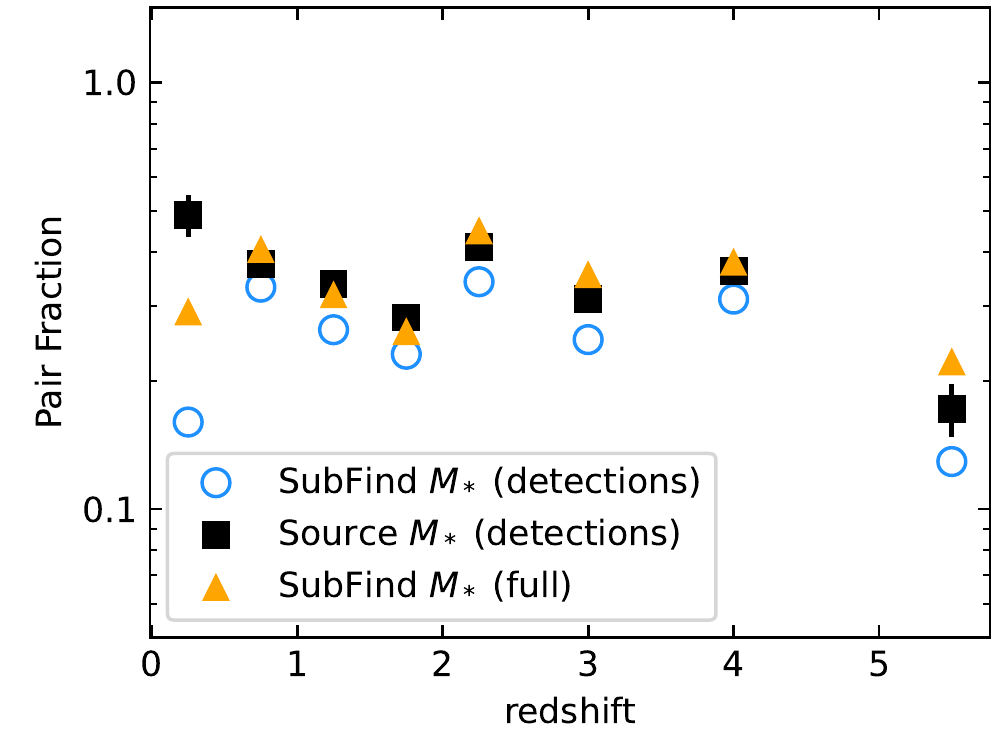}
\caption{Comparison of close pair selections from the TNG100 simulation, based on mass ratios defined in three different ways: from the SubFind catalogs alone, from the F200W source detections and stellar mass maps alone, and from the SubFind catalogs only for detected sources. We find that the pair fractions and their evolution are similar in all three cases, confirming previous results that the Illustris-derived merger pair fractions are flat or declining with higher redshift at $z \gtrsim 1$ \citep{Snyder2017,Mantha2018,Duncan2019,Pena2021}. \label{fig:pairselection}}
\end{center}
\end{figure}

\section{Image Analysis Examples} \label{s:analysis}

In this section, we present a few example science analyses based the mock data products, including close pair statistics and galaxy morphologies. We focus on one of our two $137$ arcmin$^2$ fields, the TNG100-7-6-xyz field. We create simple data simulations from the F200W images, using a simple PSF model (2D normal distribution) and adding normally distributed random shot noise to the image. For the PSF, we use a FWHM of $0.08$ arcsec.  For the noise, we use a level such that the residual background noise has $\sigma = 0.02$ nanoJanskies, which corresponds to very deep observations.

We use the resulting F200W image to detect and deblend sources, and to measure basic quantities, using the PhotUtils package \citep{Bradley2021}.  We also measure an image-based stellar mass for each source, by applying these detections to the stellar mass images we created for each field: We sum all pixels from the mass map within each source's F200W-based segmentation map.

\subsection{Close Pairs}

We follow \citet{Snyder2017} and \citet{Pena2021} to analyze the close pair statistics in our mock catalogs and images. These previous works measured close pair fractions versus redshift based exclusively on the simulations' subhalo catalogs, finding that the pair fraction evolution flattens or even declines at $z > 1$ for both the original Illustris and IllustrisTNG simulations. However, \citet{Rodriguez-Gomez2015} had identified a potential issue with this approach: it can be difficult to correctly assign mass to the primary and secondary subhalos during a merger, leading to incorrect mass ratio definitions for close pairs based only on the subhalo catalog-based masses. 

We define major close pairs as having the following properties:
\begin{itemize}
    \item Mass ratio $M_2/M_1 \ge 0.25$
    \item Separation $ 5$kpc $ < d < 70$ kpc
    \item Redshifts $\Delta z < 0.02 (1+z_{\rm 1})$ 
\end{itemize}

In Figure~\ref{fig:pairselection}, we contrast the pure subhalo catalog-based definition with a mass ratio based on source detection and image analysis for the TNG100 simulation. We test three mass ratios definitions:
\begin{enumerate}
    \item Based on the entire subhalo (SubFind) catalogs alone as was done by \citet{Snyder2017},
    \item Based on the detections applied to the stellar mass images, and
    \item Based on the subhalo catalogs for only detected sources.
\end{enumerate}
We find that all three definitions lead to broadly similar, relatively flat or declining pair fraction measurements as redshift increases. Thus, we conclude that the Illustris and IllustrisTNG-based simulations robustly predict a flat or declining pair fraction estimates in distant galaxies. This implies that in order to recover the intrinsic merger rate decline over cosmic time \citep{Rodriguez-Gomez2015}, the observability time of merging pairs must decrease with higher redshift at $z \gtrsim 1$.

\subsection{Galaxy Morphology}

We measure parametric and non-parametric galaxy morphology statistics from the 137 arcmin$^2$ mock F200W image using the StatMorph code \citep{Rodriguez-Gomez2019}\footnote{https://github.com/vrodgom/statmorph/releases/tag/v0.4.0}. In this paper, we use the Gini and $M_{20}$ statistics \citep{Lotz2004} to characterize the main morphological type of the simulated galaxies.

\begin{figure}
\begin{center}
\includegraphics[width=3.0in]{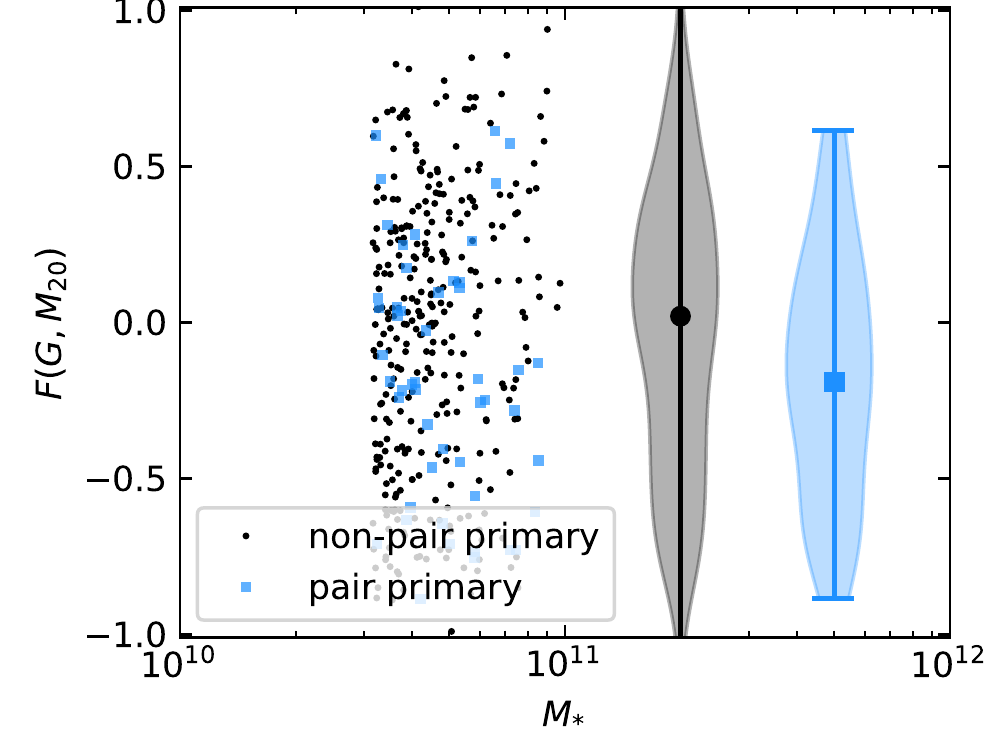}
\caption{Identification of similar morphological properties between simulated galaxies in pairs versus isolated: massive galaxies with a major companion may be slightly more disk-dominated than isolated ones. The points show the F200W morphology versus mass of the pair and non-pair samples at $0.5 < z < 4$, while the violin plots show the distribution of morphology in each sample, placed at arbitrary offset in the x direction. The points on the violin plots show the sample medians, 0.02 for isolated and -0.19 for paired. \label{fig:morphpairs}}
\end{center}
\end{figure}

In Figure~\ref{fig:morphpairs}, we separate massive galaxies ($10^{10.5} \le M_{*}/M_{\odot} \le 10^{11}$) at $0.5 < z < 4$ into pair and non-pair samples, and show the distribution of their structure as measured by the $F(G,M_{20})$ bulge statistic \citep{Snyder2015}:
\begin{equation}
F(G,M_{20}) = -0.693 M_{20} + 4.95 G - 3.96
\end{equation}
This statistic measures the position of a source along a locus in the Gini-$M_{20}$ plane, where bulge-dominated galaxies have $F \gtrsim 0.0$ and disk-dominated galaxies have $F \lesssim 0.0$. It correlates tightly with other measurements of bulge strength, such as Sersic $n$ and Concentration \citep{Bershady2000,Conselice2003a}. We find that IllustrisTNG massive galaxies with a close major companion tend to have similar, but slightly diskier, morphologies according to this metric.

\begin{figure}
\begin{center}
\includegraphics[width=3.0in]{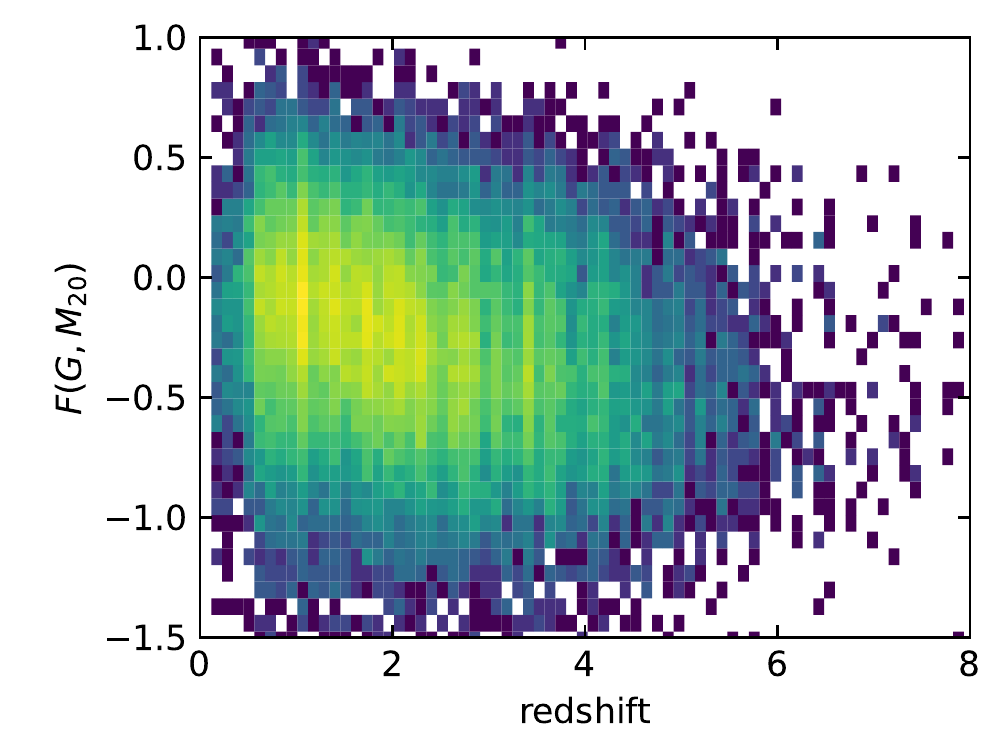}\\
\includegraphics[width=3.0in]{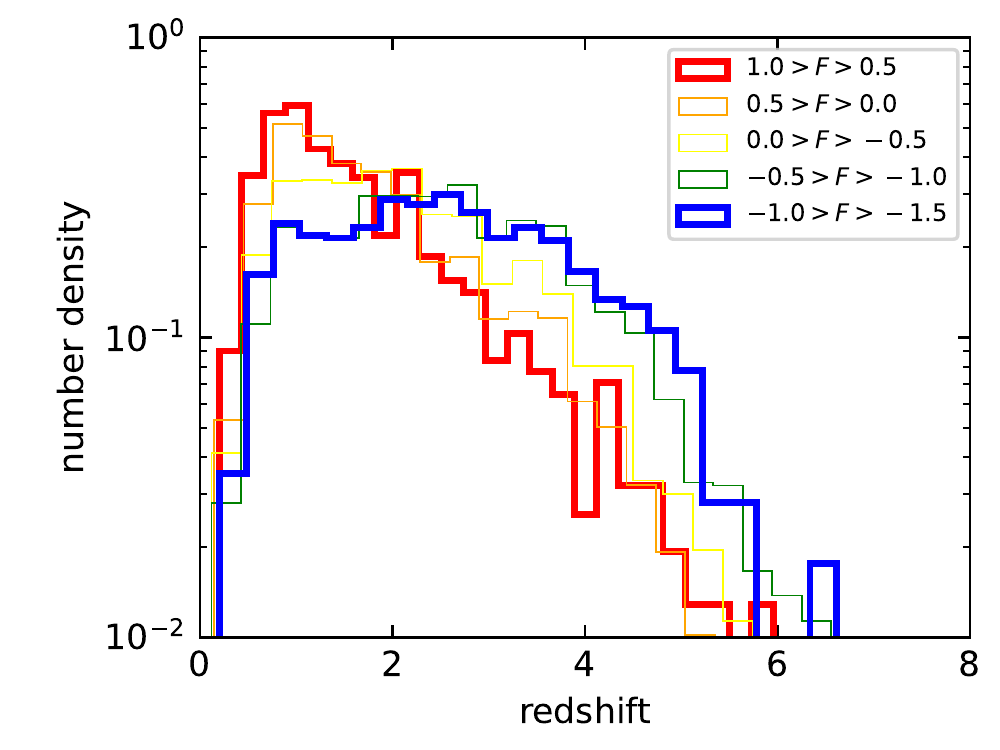}
\caption{ Top: Evolution of $F(G,M_{20})$ bulge statistic versus redshift in a mock $137$ arcmin$^2$ F200W image from the TNG100-1 simulation. This plot shows a log scaled number density.  Bottom: For selected ranges of $F(G,M_{20})$, this panel shows number density versus redshift, highlighting the fact that important numbers of bulge-dominated sources (larger $F$) emerge at $z \lesssim 3$ . \label{fig:morphs}}
\end{center}
\end{figure}

The first panel of Figure~\ref{fig:morphs} shows the distribution of $F(G,M_{20})$ as a function of redshift. We find that a significant population of bulge-dominated galaxies ($F > 0$) emerges after about $z \sim 3$. The second panel shows the redshift distribution for various ranges of $F$, where the distribution of bulge-dominated galaxies is skewed toward lower redshift than disk-dominated galaxies.

\begin{figure*}
\begin{center}
\includegraphics[width=6.5in]{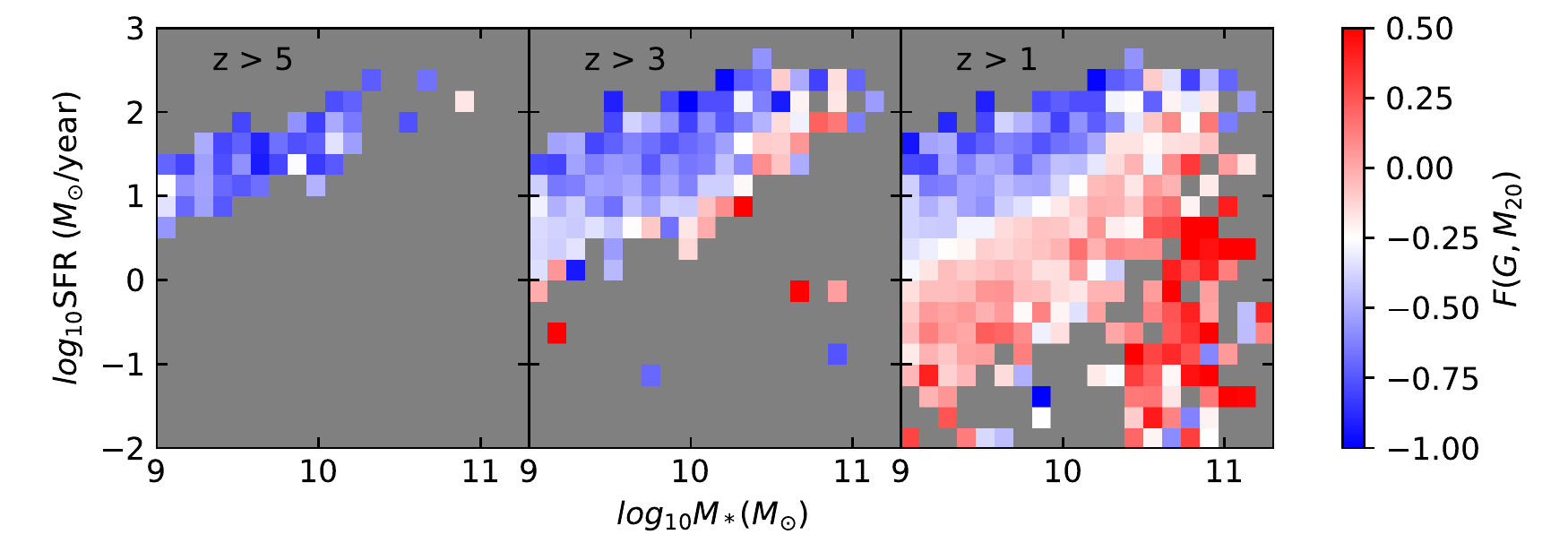}
\caption{ SFR versus $M_*$ at different redshift ranges in a mock 137 arcmin$^2$ F200W image from the TNG100-1 simulation, with bin color indicating the median value of $F(G,M_{20})$ in each bin.   \label{fig:sfr_mstar}}
\end{center}
\end{figure*}

Figure~\ref{fig:sfr_mstar} shows the evolution of star formation rate (SFR) versus stellar mass ($M_*$), color-coded by morphology using the same bulge statistic $F(G,M_{20})$. At $z > 5$, almost all galaxies lie on a narrow $SFR-M_*$ locus, and predominately have $F \lesssim -0.5$ with a few $F \gtrsim 0$ sources mixed in. For $z > 3$, a greater proportion of sources have $F \gtrsim 0.0$ and there are very few sources that start to fall below the primary $SFR-M_*$ locus.  For $z > 1$, the $SFR-M_*$ locus is much wider, with $F > 0$ sources occupying the lower portion and $F \lesssim -0.5$ sources occupying the upper portion.  There are now many more (mostly bulge-dominated) sources at high mass that fall below the main $SFR-M_*$ locus.

\section{Conclusions}

We have developed mock extragalactic survey fields derived from the IllustrisTNG simulations. We showed a few examples of how these data products could be used to inform our interpretation of galaxy evolution over cosmic time. There are many other possible applications, including but not limited to testing analysis algorithms, merger rate measurements, and deep morphological studies. We have confirmed merger pair studies based on the Illustris and IllustrisTNG simulations that imply the merger pair observability time decreases as redshift increases at $z \gtrsim 1$.  We found that massive galaxies in a major pair tend to be slightly diskier than their isolated counterparts. Finally, we show that in the TNG100 simulation, sources shaped like bulges become much more common at $z \lesssim 3$.

Mock data like these will be valuable for analyzing real JWST data from a variety of planned and future surveys, including but not limited to COSMOS-Web, the Cosmic Evolution Early Release Science survey (CEERS), the JWST Advanced Deep Extraglactic Survey (JADES), Public Release IMaging for Extragalactic Research (PRIMER), and the Next Generation Deep Extragalactic Exploratory Public (NGDEEP) survey.

\section*{Acknowledgements}

GS acknowledges support from the CEERS ERS program and the STScI Director's Discretionary research funds. AY is supported by an appointment to the NASA Postdoctoral Program (NPP) at NASA Goddard Space Flight Center, administered by Oak Ridge Associated Universities under contract with NASA. This research made use of Astropy,\footnote{http://www.astropy.org} a community-developed core Python package for Astronomy \citep{astropy2013, astropy2018}, and Matplotlib \citep{Hunter:2007}. This research made use of Photutils, an Astropy package for detection and photometry of astronomical sources \citep{Bradley2021}.

\section*{Data Availability}

The lightcone catalogs and pristine mock images are made publicly available as a MAST HLSP, with DOI \href{https://doi.org/10.17909/T98385}{https://doi.org/10.17909/T98385} and URL \href{https://archive.stsci.edu/hlsp/illustris}{https://archive.stsci.edu/hlsp/illustris}.



\bibliographystyle{mnras}
\bibliography{library} 




\appendix

\section{Description of Mock Data Products} \label{app:products}

The initial lightcone catalogs are stored in ascii text format.  The header of these files includes some basic information about how the file was created, including the name of the simulation from which it was derived (for example, TNG100-1 in the case of the ``tng100-7-6'' lightcones). Table~\ref{tab:catalogs} describes each column of the lightcone catalogs.

The mock images are stored in multi-extension FITS files.  In addition, each image FITS file contains a copy of the lightcone catalog from which it was derived, and prepends seven new columns reporting the status and properties of each subhalo's cutout that was added to the image.  

These data products are made publicly available as a MAST HLSP, with DOI \href{https://doi.org/10.17909/T98385}{https://doi.org/10.17909/T98385} and URL \href{https://archive.stsci.edu/hlsp/illustris}{https://archive.stsci.edu/hlsp/illustris}.

\begin{table*}
\centering
\caption{Lightcone catalog column definitions, for both the initial text files and FITS binary tables. }
\label{tab:catalogs}
\begin{tabular}{ccccc}
Lightcone col. no. & Image col. no. & Column Name & Column Description & units\\
\hline
N/A & 1 & \verb image_success & Whether or not this subhalo was added to the mock image & \\
N/A & 2 & \verb primary_flag & Whether or not this subhalo is a primary subhalo & \\
N/A & 3 & \verb photrad_kpc & photometric size of the subhalo & kpc \\
N/A & 4 & \verb cutoutfov_kpc & size of the cutout for this subhalo & kpc\\
N/A & 5 & \verb cutout_size & size of cutout in pixels & \\
N/A & 6 & \verb n_arcmin & size of cutout in arcmin & arcmin\\
N/A & 7 & \verb total_quant & sum of the image quantity in cutout & image units\\

\hline
1 & 8 & \verb Snapshot  \verb number & The snapshot from which this source was derived  & \\
2 & 9 & \verb Subhalo  \verb index & The unique identifier for this subhalo from this snapshot  & \\
3 & 10 & \verb RA  \verb degree & right ascension & degrees \\
4 & 11 & \verb DEC  \verb degree & declination  & degrees\\
5 & 12 & \verb RA  \verb true  \verb z & position in image plane at true z  & kpc \\
6 & 13 & \verb DEC  \verb true  \verb z & position in image plane at true z & kpc \\
7 & 14 & \verb RA  \verb inferred  \verb z & position in image plane at inferred z  & kpc\\
8 & 15 & \verb DEC  \verb inferred  \verb z & position in image plane at inferred z & kpc\\
9 & 16 & \verb True  \verb z & True cosmological redshift & \\
10 & 17 & \verb Inferred  \verb z & Inferred redshift (includes peculiar v) & \\
11 & 18 & \verb Peculiar  \verb z & Peculiar redshift  & \\
12 & 19 & \verb True  \verb scale & True scale at cosmological z  & kpc/arcsec\\
13 & 20 & \verb Comoving  \verb X & Comoving X in Observer Coordinates & Mpc\\
14 & 21 & \verb Comoving  \verb Y & Comoving Y in Observer Coordinates &  Mpc\\
15 & 22 & \verb Comoving  \verb Z & Comoving Z in Observer Coordinates & Mpc\\
16 & 23 & \verb True  \verb angular  \verb distance & True angular diameter distance to observer & Mpc\\
17 & 24 & \verb Inferred  \verb angular  \verb distance & Inferred Angular Diameter Distance to observer  & Mpc\\
18 & 25 & \verb Snapshot  \verb z & Snapshot redshift  & \\
19 & 26 & \verb Geometric  \verb z & Redshift at center of this cylinder  & \\
20 & 27 & \verb Lightcone  \verb number & Lightcone cylinder number  & \\
21 & 28 & \verb Stellar  \verb mass  \verb w2sr & Stellar mass within 2X stellar half mass radius & $M_{\odot}$\\
22 & 29 & \verb Total  \verb gas  \verb mass  \verb w2sr & Total gas mass within 2X stellar half mass radius & $M_{\odot}$\\
23 & 30 & \verb Total  \verb subhalo  \verb mass & Total mass of this subhalo (excludes children subhalos) &  $M_{\odot}$\\
24 & 31 & \verb Total  \verb BH  \verb mass  \verb w2sr & Total BH mass within 2X stellar half mass radius & $M_{\odot}$\\
25 & 32 & \verb Total  \verb baryon  \verb mass  \verb w2sr & Total baryon mass within 2X stellar half mass radius & $M_{\odot}$\\
26 & 33 & \verb SFR  \verb w2sr & SFR within 2X stellar half mass radius & $M_{\odot}$\\
27 & 34 & \verb Total  \verb BH  \verb accretion  \verb rate & Total BH accretion rate within subhalo & $\frac{10^{10} M_{\odot}/h}{0.978 Gyr/h}$\\
28 & 35 & \verb Camera  \verb X & Camera X in Observer Coordinates (Proper X at z)  & Mpc\\
29 & 36 & \verb Camera  \verb Y & Camera Y in Observer Coordinates (Proper Y at z) & Mpc\\
30 & 37 & \verb Camera  \verb Z & Camera Z in Observer Coordinates (Proper Z at z) & Mpc\\
31 & 38 & \verb Intrinsic  \verb g  \verb mag & Intrinsic stellar g absolute magnitude (BC03)  & AB mag\\
32 & 39 & \verb Intrinsic  \verb r  \verb mag & Intrinsic stellar r absolute magnitude (BC03) &  AB mag\\
33 & 40 & \verb Intrinsic  \verb i  \verb mag & Intrinsic stellar i absolute magnitude (BC03) &  AB mag\\
34 & 41 & \verb Intrinsic  \verb z  \verb mag & Intrinsic stellar z absolute magnitude (BC03) &  AB mag\\
35 & 42 & \verb Galaxy  \verb motion  \verb X & Motion in transverse Camera X direction  & km/s\\
36 & 43 & \verb Galaxy  \verb motion  \verb Y & Motion in transverse Camera Y direction  & km/s\\
37 & 44 & \verb Galaxy  \verb motion  \verb Z/Peculiar & Motion in line-of-sight Camera Z direction ; the Peculiar Velocity & km/s \\
38 & 45 & \verb Cosmological  \verb expansion &  Cosmological expansion velocity (Column 10 measures 37+38)& km/s\\
39 & 46 & \verb Apparent  \verb total  \verb gmag & Apparent total rest-frame g-band magnitude (BC03) & AB mag \\
\hline
\end{tabular}
\end{table*}


\bsp	
\label{lastpage}
\end{document}